\documentclass[aps,prd,twocolumn,nofootinbib,superscriptaddress,preprintnumbers]{revtex4-2}

\usepackage{dcolumn}
\usepackage[utf8]{inputenc}
\usepackage{amsmath,amssymb}
\usepackage{bm}
\usepackage{CJK}
\usepackage{color}
\usepackage{makecell}
\usepackage{dsfont} 
\usepackage{graphicx}
\usepackage{textcomp}
\usepackage{multirow}
\usepackage{subfiles}
\usepackage{tikz}
\usepackage[pdftex,colorlinks=true,
    linkcolor=blue,
    filecolor=blue,
    urlcolor=blue,
    citecolor=blue,
    plainpages=false,
    bookmarksopen=true]{hyperref} 

\newcommand{\getMITAffiliation}{\affiliation{Center for Theoretical Physics, Massachusetts Institute of Technology, Cambridge, Massachusetts 02139, USA}}
\newcommand{\getIAIFIAffiliation}{\affiliation{The NSF AI Institute for Artificial Intelligence and Fundamental Interactions}}

\DeclareMathOperator{\Tr}{Tr}

\begin{document}
\begin{CJK*}{UTF8}{bsmi}
\title{Neural-network preconditioners for solving the Dirac equation in lattice gauge theory}

\author{Salvatore~Cal{\`i}}
\getMITAffiliation
\author{Daniel~C.~Hackett}
\getMITAffiliation
\getIAIFIAffiliation
\author{{Yin~Lin} (林胤)}
\getMITAffiliation
\getIAIFIAffiliation
\author{Phiala~E.~Shanahan}
\getMITAffiliation
\getIAIFIAffiliation
\author{Brian~Xiao}
\getMITAffiliation

\date{\today}

\preprint{MIT-CTP/5449}

\begin{abstract}
This work develops neural-network--based preconditioners to accelerate solution of the Wilson-Dirac normal equation in lattice quantum field theories. 
The approach is implemented for the two-flavor lattice Schwinger model near the critical point. In this system, neural-network preconditioners are found to accelerate the convergence of the conjugate gradient solver compared with the solution of unpreconditioned systems or those preconditioned with conventional approaches based on even-odd or incomplete Cholesky decompositions, as measured by reductions in the number of iterations and/or complex operations required for convergence.
It is also shown that a preconditioner trained on ensembles with small lattice volumes can be used to construct preconditioners for ensembles with many times larger lattice volumes, with minimal degradation of performance.
This volume-transferring technique amortizes the training cost and presents a pathway towards scaling such preconditioners to lattice field theory calculations with larger lattice volumes and in four dimensions.
\end{abstract}

\maketitle
\end{CJK*}

\section{Introduction}

Lattice quantum field theory (LQFT) is a non-perturbative regularization of quantum field theory that enables numerical calculations in the strong coupling regime. For example, LQFT is the only \textit{ab initio} approach to calculating hadronic observables from quantum chromodynamics (QCD) and has enabled significant contributions to our understanding of non-perturbative processes in the Standard model; see Ref.~\cite{ParticleDataGroup:2020ssz,Detmold:2019ghl,Bazavov:2019lgz,USQCD:2019hyg,Kronfeld:2019nfb,Cirigliano:2019jig,USQCD:2019hee,Joo:2019byq} for recent reviews.

In many LQFT calculations with fermions, the dominant computational cost arises from solving systems of linear equations
\begin{equation}
    Ax = b
\label{eq:first}
\end{equation}
for square, sparse matrices $A$ constructed from lattice Dirac operators. 
These systems arise both in the generation of gauge field configurations, and in the computation of fermionic observables. In either case, the linear systems must be solved with $\sim$$100$s of right-hand sides $b$ and $\sim$$1000$s of different matrices $A$, with typical matrix sizes on the order of $10^6$ to $10^9$ on each dimension.

Most LQFT calculations use iterative solvers, typically Krylov subspace methods like the conjugate gradient (CG) algorithm~\cite{booksaad}, to solve Eq.~\eqref{eq:first}. 
These methods---which iteratively construct approximate solutions until some target accuracy is achieved---provide an efficient approach to solving systems of linear equations at scales 
which would be intractable using direct methods such as Gaussian elimination or $LU$ decomposition. Their computational cost, however, is still significant, particularly in the context of lattice QCD calculations with small physical lattice spacings and light quark masses, where 
the number of iterations required for convergence becomes large due to increasingly large condition numbers~\cite{Joo:2019byq}.  
Accelerating these algorithms would enable more precise calculations given fixed computing resources.
 
One approach to accelerating the solution of linear equations is via preconditioning, i.e., the transformation of a linear system into a different but equivalent linear system that is better conditioned, and thus easier to solve in terms of time-to-solution. The quality of any preconditioning procedure is a result of the inherent trade-offs between the costs of constructing and applying the preconditioner and its ability to accelerate convergence.
Some of the most commonly used preconditioners in LQFT calculations include even-odd preconditioners~\cite{DeGrand:1988vx}, inexact deflation~\cite{Luscher:2007se}, algebraic multigrid methods~\cite{Brannick:2007ue,Babich:2010qb,Osborn:2010mb,Frommer:2013fsa,Brannick:2014vda,Brower:2018ymy,Brower:2020xmc}, and the Schwarz alternating procedure~\cite{Luscher:2003qa}. There have also been attempts to apply other preconditioners, such as incomplete $LU$ decomposition~\cite{Oyanagi:1986dm} and symmetric successive over-relaxation~\cite{Fischer:1996th}, to LQFT systems. 
Different preconditioners can often complement one another and may be used in sequence; it is thus of great interest to continue to explore new approaches to precondition Dirac equations.

In this work, we present a framework for constructing preconditioners for 
Dirac normal matrices using convolutional neural networks (CNNs) with sparse~\cite{sparse} and dense convolutions, and apply the approach to the unquenched lattice Schwinger model in two dimensions~\cite{Schwinger:1962tp} as a precursor and proof-of-concept for future applications to other lattice field theories such as lattice QCD in four dimensions. We demonstrate that neural-network preconditioners are able to accelerate the convergence of CG solves in this context. Similar network architectures have been used to construct preconditioners for solving linear equations for fluid simulation and computer vision~\cite{waterpaper,li2020learning}, with some success in achieving higher efficiency than other algorithms such as algebraic multigrid. 

To train the preconditioner models, we use a single ensemble of $\mathrm{U}(1)$ gauge fields for a theory with Wilson fermions~\cite{Wilson:1974sk}, two degenerate sea quarks, nearly critical parameters, and lattice volume $32^2$. After optimizing the network parameters,
we find that the number of CG iterations required for convergence of preconditioned Dirac matrices is reduced by a factor of between two and five over unpreconditioned solves.
However, when assessing computational advantage, the costs of applying the preconditioners must also be considered. In particular, we find that preconditioners constructed with only sparse convolutions result in the best performance in terms of the number of complex operations to solution, even though the number of iterations to convergence for these constructions is larger than that resulting from other preconditioners.

The preconditioner architectures based on CNNs which are developed here are agnostic to the lattice volume,
so a network trained on one lattice ensemble can trivially produce preconditioners for other volumes.
We find that a network trained on an ensemble of lattice volume $8^2$ produces preconditioners for larger lattice volumes that are just as effective as preconditioners trained directly on larger-volume ensembles. 
This volume-transferring technique provides an efficient method to optimize preconditioners for large lattice volumes and will be important for future applications of this approach to lattice QCD calculations, for which the computational and memory costs of training on typical lattice volumes would likely be prohibitive with current approaches and hardware.

\section{Preconditioning lattice Dirac normal equations}

This section introduces the numerical problem of solving Dirac equations in a LQFT and outlines how preconditioning techniques can help to accelerate solver convergence. 

Calculating fermionic observables in a lattice gauge theory requires solving the linear equation
\begin{align}
    D_{\alpha x,\beta y}\psi_{\beta y} = \eta_{\alpha x}
    \label{eq:dirac_eq}
\end{align}
to obtain the quark propagator $\psi$ for a given lattice Dirac matrix $D$ and source vector $\eta$. Here, the subscripts $\alpha$ and $\beta$ denote the combined spin and color degrees of freedom in the theory, and $x$ and $y$ denote the sites of the lattice. All repeated symbols are implicitly summed over.
In this work, we consider solving the normal equation resulting from left-multiplying the Dirac equation with $D^\dagger$, such that the problem is of the form of Eq.~\eqref{eq:first} with
\begin{equation}
    A \equiv D^\dagger D, ~x \equiv \psi,~\text{and}~b \equiv D^\dagger x,
    \label{eq:normal_def}
\end{equation}
where $A$ is a Hermitian, positive-definite (HPD) matrix, and all indices have been suppressed.

 The $n= (V\times d)$-dimensional vector space of the matrix $A$ spans both the spacetime and internal degrees of freedom, where $V$ is the lattice volume and $d$ is the total dimension of the spinor and color degrees of freedom. 
 Although often large, the Dirac matrix is typically highly sparse, with the number of non-zero entries approximately proportional to $n$.

To precondition Eq.~\eqref{eq:first}, let $M_L^{-1}$ and $M_R^{-1}$ be the so-called left and right preconditioners, which are non-singular, square matrices of the same size as $A$. The preconditioned system is defined to be
\begin{align}
    A^\prime x^\prime = b^\prime,
    \label{eq:precond}
\end{align}
where $A^{\prime} \equiv M_L^{-1}AM_R^{-1}$, $x^\prime\equiv M_{R}x$, and $b^\prime \equiv M_L^{-1}b$. Equivalently, the preconditioned CG algorithm~\cite{booksaad} can be used to solve the unpreconditioned system, which has the same effect as solving the preconditioned system with the standard CG algorithm but avoids the complication of computing and multiplying explicit representations of $M_L^{-1}$ and $M_R^{-1}$.

The rate of convergence of iterative solvers is governed by the condition number of the matrix $A$ or ($A'$)~\cite{booksaad}:
\begin{align}\label{eq:conditionnumber}
    \kappa(A) \equiv \frac{\sigma_\text{max}}{\sigma_\text{min}} = \frac{|\lambda_\text{max}|}{|\lambda_\text{min}|} ~ ,
\end{align}
where $\sigma_\text{min}$ and $\sigma_\text{max}$ are the smallest and largest singular values of $A$, respectively, which are equal to the absolute values of the corresponding eigenvalues $|\lambda_\text{min}|$ and $|\lambda_\text{max}|$ since $A$ is Hermitian.
Preconditioning attempts to alleviate the numerical problem by producing a better-conditioned system with $\kappa(A^\prime) \ll \kappa(A)$. 
An ideal preconditioner needs to achieve a balance between how close $A^\prime$ is to the identity by some metric---and hence the number of CG iterations required to solve the preconditioned system---and how costly it is to numerically construct and apply the preconditioner.

In this work, the even-odd and incomplete Cholesky (IC) preconditioners are used as baselines against which the performance of neural-network preconditioners is measured.
We avoid the complexity of a comparison with algebraic multigrid methods, for which a fair comparison would require exploration of the broad family of different possible implementations on each lattice ensemble.
The details of even-odd and IC preconditioners are outlined in App.~\ref{appdx:precond}.

\section{Neural-network preconditioners for the lattice Schwinger Model}

This section describes the construction of neural-network preconditioners for solving Dirac normal equations in the lattice Schwinger model.

\subsection{The lattice Schwinger model}
The lattice action of the two-flavor Schwinger model can be defined as the standard plaquette action with two degenerate Wilson fermions:
\begin{align}
\begin{split}
    S = &-\beta\sum_{x}\text{Re}\big(P_x\big) +
    \sum_{f=0}^1\sum_{x,y} \overline{\psi}^{(f)}_x D_{x,y}\psi^{(f)}_y, 
\end{split}
\end{align}
where
\begin{equation}
    P_x =  U_{1,x}U_{2,x+\hat{1}}U^*_{1,x+\hat{2}}U^*_{2,x}
\end{equation}
is the plaquette, and
\begin{align}
    D_{x,y} &= (m+2r)\delta_{x,y} 
    -
    \nonumber\\
    &~~~~~~
    \frac{1}{2}\sum_{\mu=1}^2\bigg(
    (1-\gamma_\mu) U_{\mu,x}\delta_{x+\hat{\mu},y}
    +
    \nonumber\\
    &~~~~~~~~~~~~~~~~
    (1+\gamma_{\mu})U^*_{\mu,x-\hat{\mu}}
    \delta_{x-\hat{\mu},y}
    \bigg)
\end{align}
is the Wilson discretization of the Dirac operator~\cite{Wilson:1974sk}.
The position on a two-dimensional lattice is labeled by $x$ or $y$. $U_{\mu,x} \in \text{U}(1)$ is the complex gauge field where ${\mu\in\{1, 2\}}$ labels the spatial and temporal components and $\psi^{(f)}_x$, $\overline{\psi}^{(f)}_x$ are two-component Wilson fermion fields with flavor indices $f\in\{0,1\}$. $\gamma_1$ and $\gamma_2$ are Euclidean gamma matrices in two dimensions. The specific representation we use here is given by Pauli matrices: $\gamma_1 = \sigma_1$ and $\gamma_2 = \sigma_2$ such that $\gamma_5 = i\gamma_1\gamma_2 = -\sigma_3$. $m$ and $\beta$ are bare lattice parameters and $r$ is the Wilson parameter; we set $r=1$ throughout this work. We also set the lattice spacing $a=1$ throughout this work.

Periodic boundary conditions are applied in all directions for the gauge field $U_{\mu,x}$; for the fermionic fields $\psi^{(f)}_x$ and $\overline{\psi}^{(f)}_x$, periodic and anti-periodic boundary conditions are applied in the spatial and temporal directions, respectively.

\subsection{Architecture}
\label{sec:arch}

\newcommand\changeme[1]{{\color{red} #1}}
\newcommand\mat[1]{{#1}}
\begin{figure*}
\includegraphics[width=1.0\textwidth]{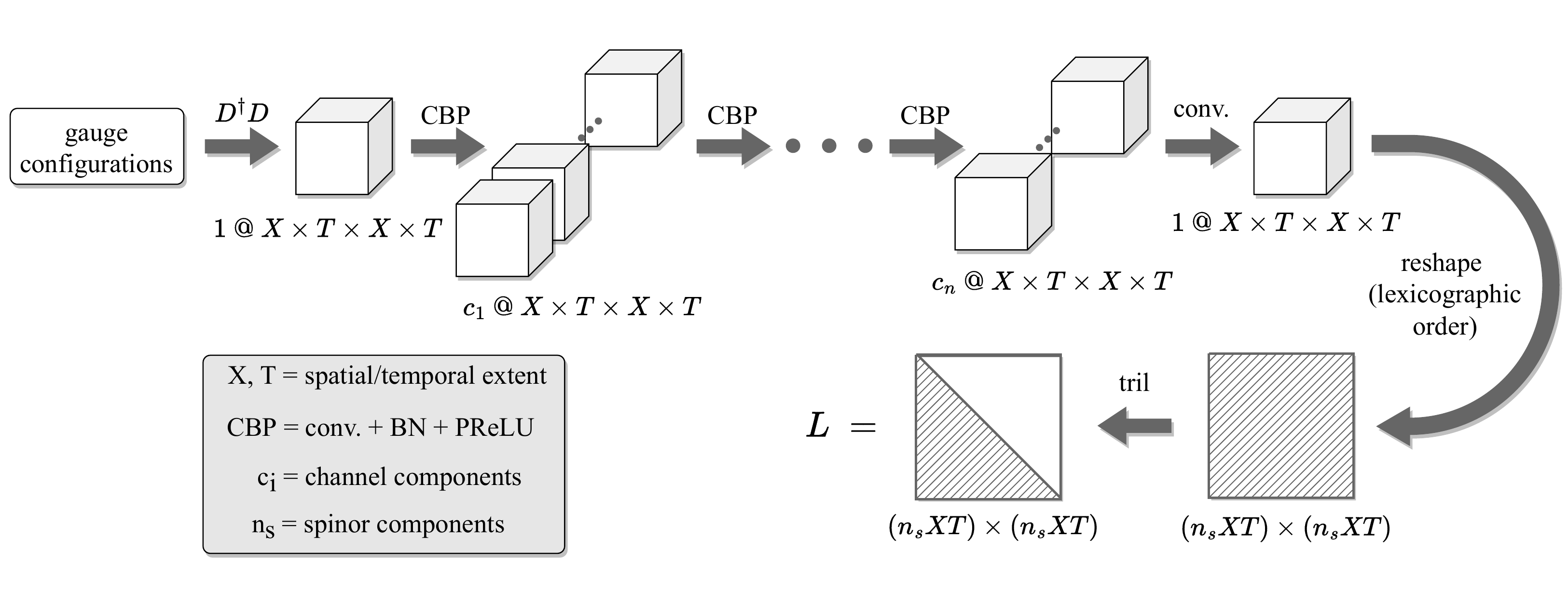}
\caption{
An illustration of the network architecture designed for preconditioning Dirac normal equations in the lattice Schwinger model. 
$c~@~X\times T\times X\times T$ denotes a tensor with $c$ channels, of dimension $(16c)\times X\times T \times X \times T$ as for each channel there are $16 = (2\times n_s)^2$ real tensor components, where $2$ is the complex dimension and $n_s=2$ is the spinor dimension.
The convolution (which is sparse or dense, depending on architecture choice as described in the text) is four-dimensional and acts on the spatial and temporal indices at the source and sink of Dirac normal matrices.
}
\label{fig:arch}
\end{figure*}
\begin{figure}
\includegraphics[width=0.48\textwidth]{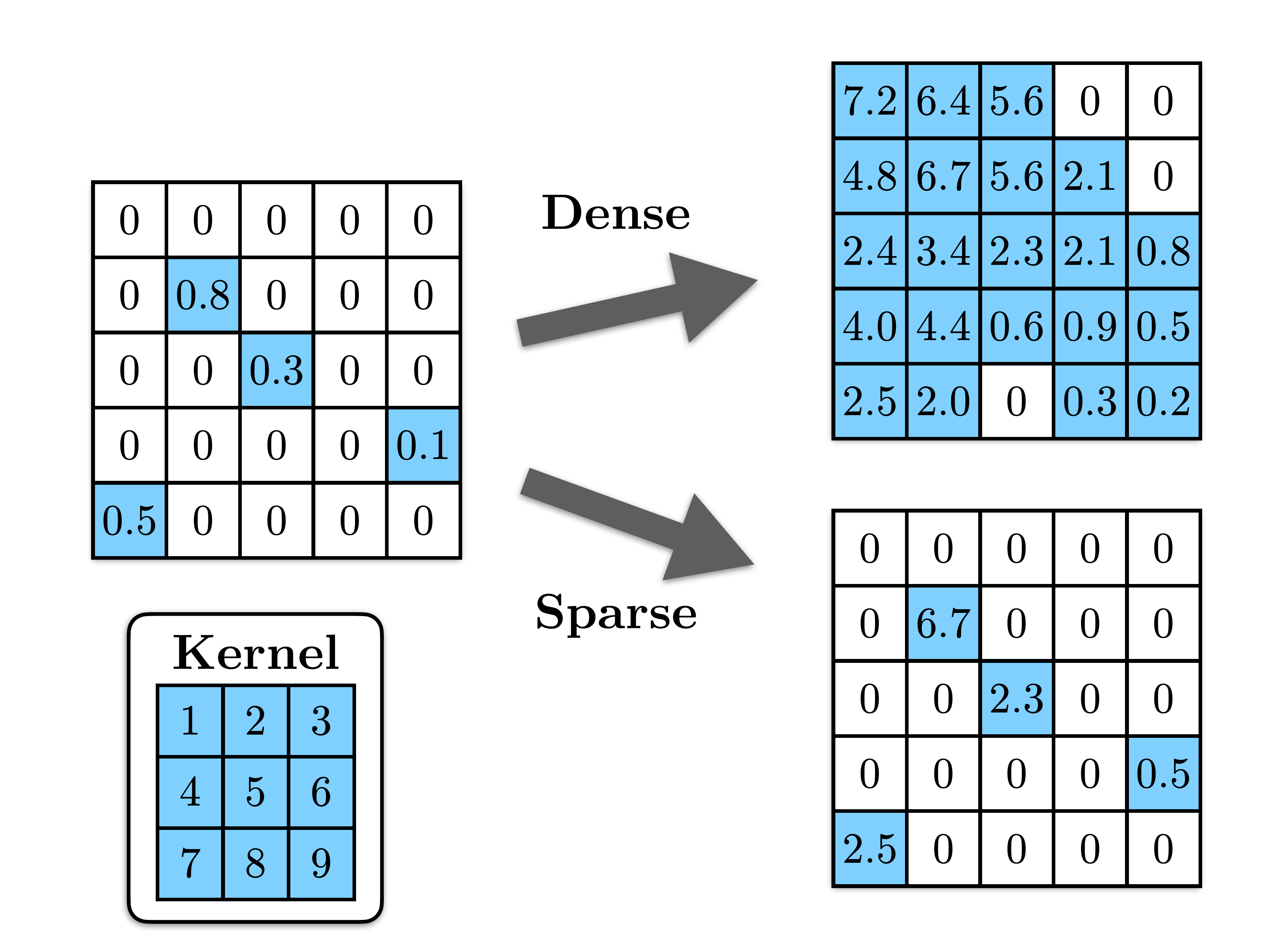}
\caption{
An illustration of the difference between the action of a conventional (dense) convolution and a sparse convolution on the same matrix. The same kernel acts on both grids, with zero padding. Non-zero entries are highlighted in blue.}
\label{fig:sparsity-conv}
\end{figure}
The architecture we construct for neural-network preconditioners, shown in Fig.~\ref{fig:arch}, has the following structure: first, a preprocessing unit prepares the Dirac normal operators to a format suitable for convolutions. This is followed by multiple convolutional units. Finally, a postprocessing unit outputs a lower triangular matrix $L$ with real and positive diagonal entries which can be used to form a preconditioner. The parameters of the neural network are optimized to minimize a loss function that is correlated with the number of CG iterations required for convergence; choices of loss functions are detailed in Sec.~\ref{sec:loss}.

\textit{Preprocessing unit} --- We explicitly construct Dirac normal matrices from gauge configurations, then reshape them into real tensors of shape $16\times X \times T\times X\times T$, where $X$ and $T$ are the spatial and temporal lattice extents, which are repeated twice to represent the source and sink indices. The first factor arises as $16=(2n_s)^2$, where $n_s = 2$ is the number of spinor components in two dimensions, and the factor of $2$ arises from the real and imaginary parts. The same tensor is denoted as $1@ X\times T\times X\times T$ in Fig.~\ref{fig:arch}, where the first index is the channel dimension, to emphasize that the convolutional kernels only convolve across spacetime dimensions. 
Note that a single channel in this work always consists of $16$ spinor and complex components.

\textit{Convolutional units} --- The main building blocks of the network are convolutional units that each consist of three layers in sequential order: the convolutional layer, batch normalization (BN) layer, and the parameterized rectifiable linear
unit (PReLU) activation layer. 

More precisely, first the input tensor is passed into a convolutional layer with a kernel size of $k_1 \times k_2 \times k_3 \times k_4$ acting on the source and sink spacetime indices. We then apply a BN layer~\cite{ioffe2015batch} to better condition the training. Finally, a PReLU layer follows every convolutional unit, except for the last unit where only the convolution is applied. A total of $n$ convolutional units, each with $c_n$ channels, are stacked together in the network.
We consider constructions with both dense and sparse convolutions---as illustrated in Fig.~\ref{fig:sparsity-conv}---for the convolutional layers to preserve the sparsity pattern of input data and constrain the costs of constructing and applying output preconditioners. 

\textit{Postprocessing unit} --- The tensor is reshaped into a two-dimensional complex matrix $B$ in the lexicographic ordering with the spinor index being the fastest-changing and the spatial index the slowest. The shape of the matrix is $(n_sXT)\times (n_sXT)$.
We then discard the upper triangular part of $B$ to yield $\mat L = \operatorname{tril} B$ and enforce the diagonal entries to be real with values equal or greater than a small parameter $\epsilon$ to ensure its invertibility.

Two types of preconditioners are constructed for each neural-network output $L$. In the \textbf{single-preconditioning method}, $M_L^{-1} = (M_R^{-1})^\dagger = L^\dagger$, whereas in the \textbf{double-preconditioning method}, $M_L^{-1} = (M_R^{-1})^\dagger = L^\dagger L$. The preconditioners will be hereafter referred to as single and double preconditioners, respectively.
The single preconditioners are inspired by the Cholesky decomposition of the Dirac normal matrices $A = L_A^\dagger L_A$ for some lower triangular matrix $L_A$ with real, positive diagonal entries.
However, as discussed below, the double preconditioners in practice produce larger reductions in the number of CG iterations required for convergence at the cost of denser preconditioners. 

\subsection{Loss function}
\label{sec:loss}
Ideally, the free parameters of the architecture should be optimized to produce preconditioners that minimize the average time needed to solve Eq.~\eqref{eq:precond} to a given precision. However, while this defines a learning task, it is not differentiable and thus not amenable to optimization with stochastic gradient descent. We instead optimize a differentiable proxy for this objective, training the networks in stages using both the condition number and the $K$-condition number as loss functions. 

The use of the condition number (Eq.~\eqref{eq:conditionnumber}) as a loss function is natural, since it pertains to the convergence theorem of the CG algorithm and other iterative solvers~\cite{booksaad}. Similarly, the $K$-condition number, which is a measure of the degree of clustering in the eigenvalue spectrum, is known to be related to the rate of convergence of the CG algorithm~\cite{Kalkreuter:1994ax,booksaad}. The $K$-condition number
of a HPD matrix $Q$ is defined~\cite{Kaporin1994NewCR}\footnote{In Ref.~\cite{Kaporin1994NewCR}, the $K$-condition number of a matrix $Q$ is denoted $B(Q)$.} as
\begin{equation}
K(Q) \equiv \frac{\frac{1}{n}\Tr(Q)}{\det(Q)^{\frac{1}{n}}} = \frac{\frac{1}{n}\sum_{i=1}^{n}\lambda_i}{\left(\prod_{i=1}^{n}\lambda_i\right)^{\frac{1}{n}}},
\label{eq:eigen_k}
\end{equation}
where $n$ is the order and $\lambda_1, \cdots, \lambda_n$ are the eigenvalues of $Q$.
From Eq.~\eqref{eq:eigen_k} we see that $K(Q) \geq 1$ and $K(Q) = 1$ if and only if $Q = \lambda I$ where $\lambda$ is the eigenvalue and $I$ is the identity matrix. 

\begin{figure}
\includegraphics[width=0.48\textwidth]{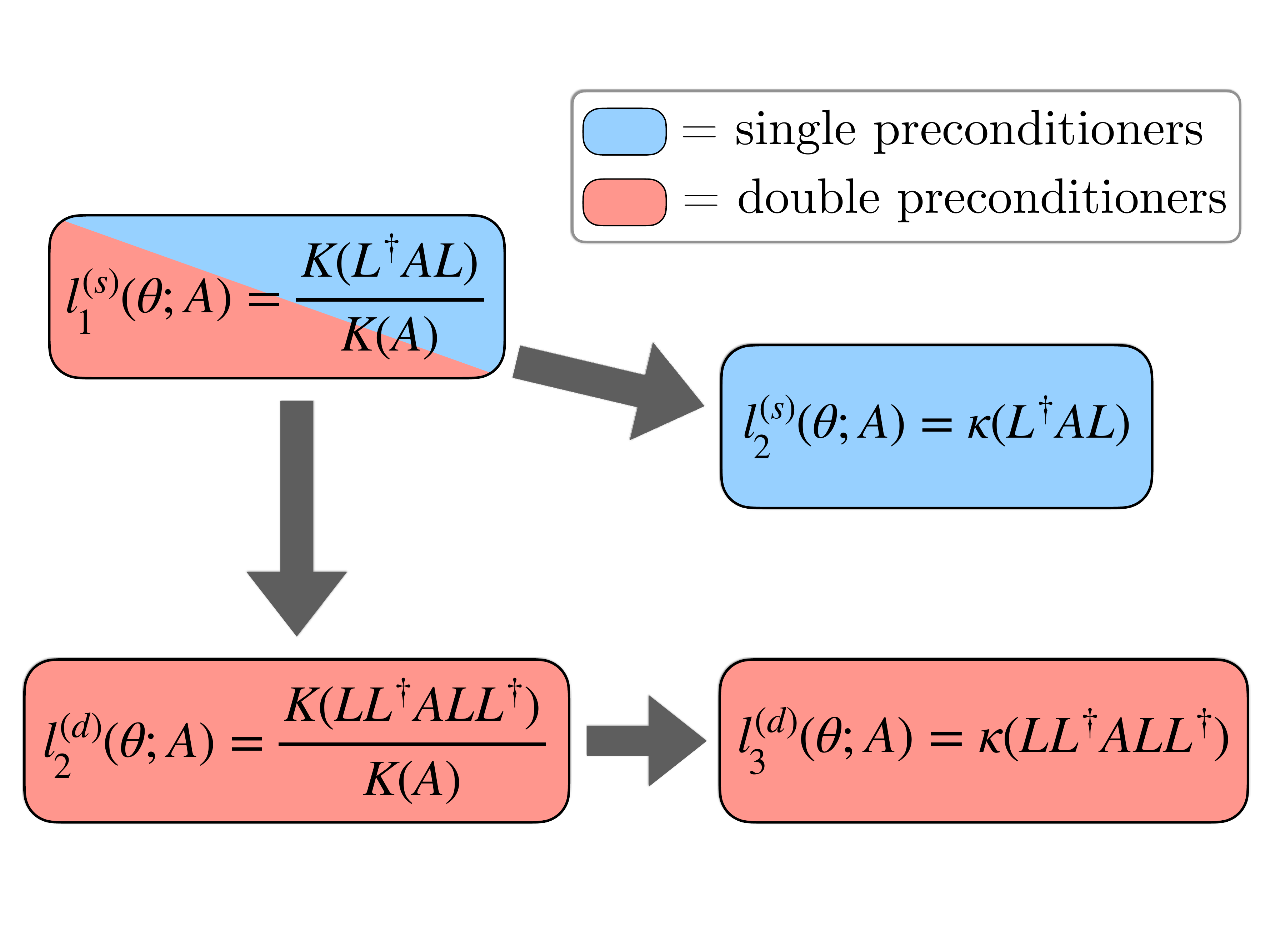}
\caption{
Loss functions for training single (blue) and double preconditioners (red) as defined in Sec.~\ref{sec:arch}. $\theta$ is the set of parameters for the neural network, on which the preconditioner $L$ implicitly depends. Both types of preconditioners are pretrained with $l_1^{(s)}(\theta; A)$.}
\label{fig:training_stages}
\end{figure}

\textit{Training single preconditioners} ---
The networks are trained in two stages. In the first stage,
the loss function that we minimize for the Dirac normal matrix $A$ is
\begin{align}
\begin{split}
    l^{(s)}_1(\theta; A) &= \frac{K(L^\dagger AL)}{K(A)}
    \\ &= \left(\frac{\frac{1}{n}\Tr(L^\dagger AL)}{\frac{1}{n}\Tr(A)}\right)
    \left(\frac{1}{\det(L^\dagger L)^{\frac{1}{n}}}\right),
\end{split}
\label{eq:l1s}
\end{align}
where $\theta$ is the set of neural network parameters which are optimized and on which the matrix $L$ implicitly depends. 
Instead of minimizing $K(L^\dagger AL)$ directly, we minimize the ratio $K(L^\dagger AL)/K(A)$ to avoid computing the determinant of the Dirac normal matrix---$\det(L^\dagger L)$ is efficient to compute since $L$ is triangular.
This means that all training gradients are weighted by an additional factor of $1/K(A)$, so Dirac normal matrices with larger $K$-condition numbers will be weighted less heavily in training. In practice, these extra normalization factors play little role, since we observe them to be almost identical across the gauge configurations in an ensemble. In the second stage of the training, we change to a new loss function, $l^{(s)}_2(\theta; A) \equiv \kappa(L^\dagger A L)$; in practice, this acts to further refine the results.

\textit{Training double preconditioners} ---
We train the network for double preconditioners in three stages.
In the first stage, we use the same loss function as the single-preconditioning method, $l^{(s)}_1(\theta; A)$. In practice, the pretrained network obtained for the single-preconditioning method can simply be re-used for this stage, without additional computation. Even though the definition of $l_1(\theta; A)$ in Eq.~\eqref{eq:l1s} is motivated by the single-preconditioning method, it is in practice an effective proxy for the number of CG iterations required for convergence for the double preconditioners. In the second and third stages of training, we refine the optimization by minimizing $l^{(d)}_2(\theta; A) \equiv K(LL^\dagger A L L^\dagger)/K(A)$ and $l^{(d)}_3(\theta; A) \equiv \kappa(LL^\dagger A LL^\dagger)$ in sequence, which are the $K$-condition number ratio and the condition number of the double preconditioned matrix.

The two strategies for training the single- and double-preconditioning methods are summarized in Fig.~\ref{fig:training_stages}. 
The motivation for the staged training approach in each case is two-fold; first, optimizing network parameters with respect to the two final loss functions $l^{(s)}_2(\theta; A)$ and $l^{(d)}_3(\theta; A)$ is computationally expensive due to the need to compute eigenvalues in the condition number. Loss functions constructed from $K$-condition numbers are less computationally expensive to evaluate, and pre-training with the $K$-condition loss functions thus reduces the overall training time in our numerical investigation. Moreover, in practice the pre-training procedure is found to improve the final results compared with those achieved with no pre-training steps.

\section{Numerical investigation}
\label{sec:results}
\setlength{\tabcolsep}{10pt}
\renewcommand{\arraystretch}{1.2}
\begin{table*}[t]
    \begin{tabular}{ccccccccc}
    \hline\hline
    $X\times T$    & $\beta$ & $\kappa$     & $ m_\pi$ & $ m_\pi X$ & $n_\text{train}$ & $n_\text{validate}$ & $n_\text{test}$ & use           \\
    \hline
    $8 \times 8$   & $2.0$   & $0.276$ &    $0.47(1)$     & $3.7(1)$&$1540$   &  $60$   &  $200$          & volume transfer \\
    $16 \times 16$ & $2.0$   & $0.276$ &        $0.21(1)$ & $3.4(2)$ & n/a & n/a    & $200$          & volume transfer \\
    $32 \times 32$ & $2.0$   & $0.276$ &     $0.12(1)$    & $3.7(3)$ & $900$ &  $60$    &    $200$       & nominal \& volume transfer         \\
    $64 \times 64$ & $2.0$   & $0.276$ &     $0.08(3)$    & $5(2)$
& n/a &  n/a  &     $32$     & volume transfer \\
    \hline\hline
    \end{tabular}
    \caption{Ensembles of the two-flavor lattice Schwinger model with Wilson fermions used for the numerical study. The $16^2$ and $64^2$ ensembles are not used for training the neural network. The ``pion'' (lightest pseudo-scalar) masses, $m_\pi$, in lattice units are obtained from one-state fits to the correlation functions and the quoted uncertainties are statistical, estimated from $200$ bootstrap samples.}
    \label{tab:ensemble}
\end{table*}

For a numerical demonstration of the training and evaluation of neural-network preconditioners we generate a set of ensembles with the HMC algorithm with parameters as detailed in Tab.~\ref{tab:ensemble}. We perform most of our tests on the ``nominal ensemble", with a $32^2$ lattice volume and $\beta=2.0$. To maximize the number of CG iterations required to converge to the solution with a given precision, i.e., to maximize the numerical difficulty of solving the unpreconditioned Dirac normal equation, we choose the hopping parameter $\kappa=1/(2(m+2)) = 0.276$ that corresponds to $m = -0.188$. 
This is close to the critical mass $m_\text{crit} \approx -0.197$ at $\beta = 2.0$~\cite{Gattringer:1997qc,Christian:2005yp}. To generate the other ensembles used in the volume transfer study, we fix the values for $\beta$ and $m$ while varying only the lattice volume.

The network architecture used is as detailed in Sec.~\ref{sec:arch}, with a kernel size of $k_1=k_2=k_3=k_4=3$, $n=3$ layers, and $c_1=c_2=c_3=12$ channels. Diagonal entries of $L$ are clamped to be equal or greater than $\epsilon = 10^{-3}$. Two types of networks are trained to investigate the effects of preconditioner sparsities on their performance. For the \textbf{sparse} networks, only sparse convolutional layers are used; on the other hand, the last two convolutional layers of the \textbf{two-dense} networks are dense and the first one is sparse. In either network, there are in total $871298$ trainable parameters.

On the $8^2$ and $32^2$ ensembles, $n_\text{train}$ configurations are used for training  and $n_\text{validate}$ configurations taken from the end of the same Monte Carlo stream are used for validation. Finally, $n_\text{test}$ configurations generated from a separate Monte Carlo stream are used for testing. We find our data are not significantly autocorrelated, checked by ensuring the uncertainties of the plaquette and topological charge mean values stay approximately constant when blocking over increasing ranges of neighboring measurements.

In all training tasks, the loss functions are minimized with a mini-batch size of $32$ gauge configurations, a learning rate of $10^{-4}$, and a gradient-clipping norm of $0.1$. We implement the network with  \texttt{PyTorch}~\cite{NEURIPS2019_9015}, together with \texttt{Minkowski Engine}~\cite{choy20194d} for the sparse convolutions. Parameters in the networks are optimized using the Adam optimizer~\cite{kingma2017adam} with \texttt{PyTorch} default parameters; the optimizer is reset for each stage of training. The network evaluations are performed in single precision to accelerate training, while the rest of computations are done in double precision. 

All networks are trained on $8$ GeForce~RTX~2080~Ti GPUs on one node with an Intel~Xeon~Gold~5218 CPU.  We sequentially minimize each one of the loss functions shown in Fig.~\ref{fig:training_stages} for $300$ epochs, so that its value evaluated on the validation dataset 
is no longer improving and varies only a few percent under additional training.
For the sparse network architecture, the training costs for the nominal ensemble with the single and double preconditioners are $210$ and $270$~RTX~2080~Ti~GPU-hours, respectively; for the two-dense network architecture, the training costs for the single and double preconditioners are $260$ and $350$~RTX~2080~Ti~GPU-hours, respectively. 
The architectures trained on the $8^2$ ensemble are optimized using the same procedure as for the nominal ensemble. 
However, the training costs are substantially less as a result of the smaller volume---for the sparse network, the training costs for the single and double preconditioners are $7$ and $10$~RTX~2080Ti~GPU-hours, respectively; for the two-dense network, the training costs for the single and double preconditioners are $12$ and $17$~RTX~2080Ti~GPU-hours, respectively.

\begin{figure}
    \centering
    \includegraphics[width=.45\textwidth]{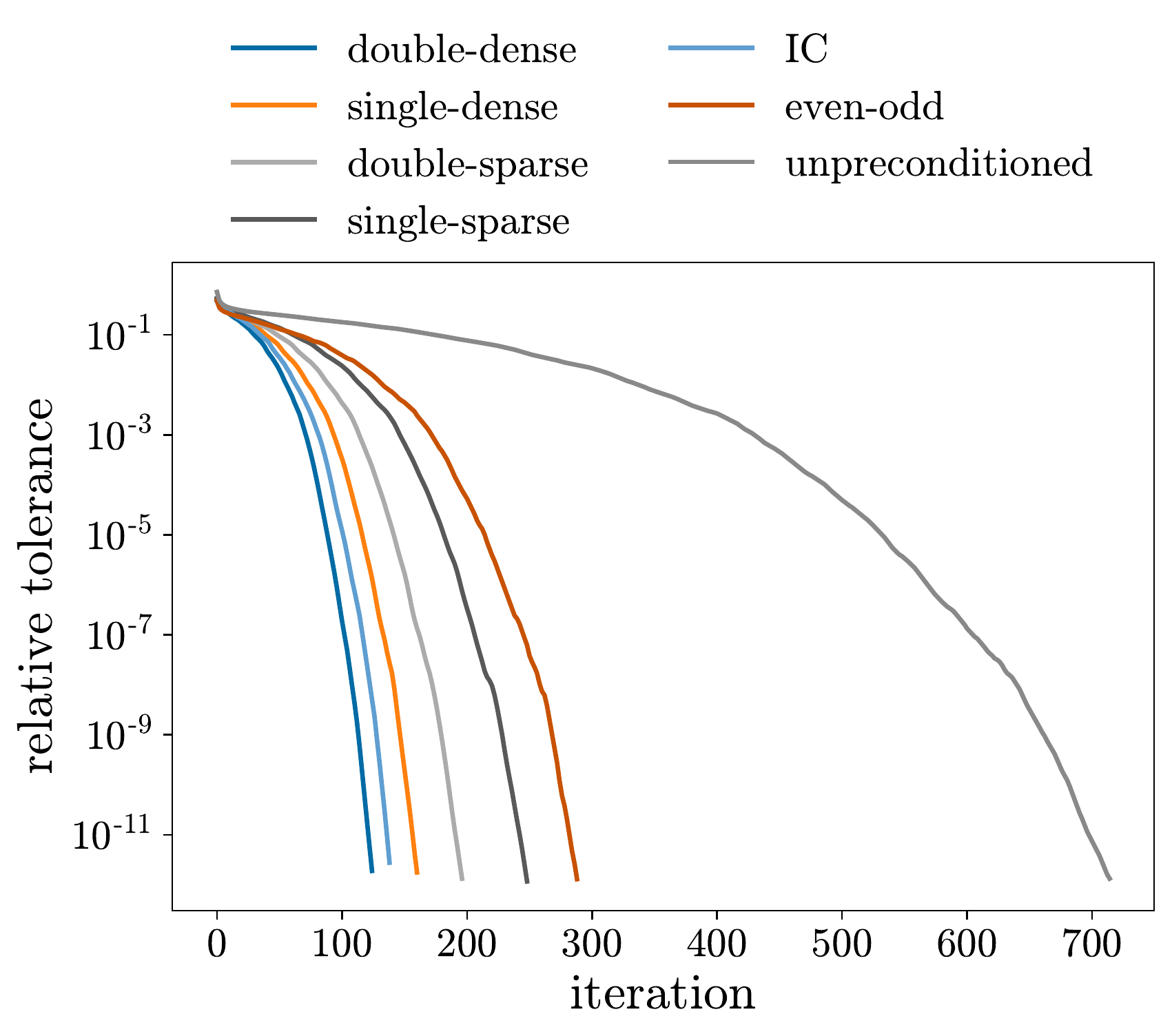}
    \caption{Relative tolerance (Eq.~\eqref{eq:rel_tol}) as a function of CG iterations for unpreconditioned and  preconditioned systems averaged over solves on $200$ test gauge configurations.
    The first and second parts of the hyphenated legend labels indicate whether the corresponding results are for the single or double preconditioning procedure, with architectures constructed from sparse or two-dense networks, respectively.}
    \label{fig:nn_residuals}
\end{figure}

\subsection{Results on the nominal ensemble}
\label{sec:nominal_results}
This section summarizes the results of the neural-network preconditioner trained on the nominal ensemble as described in the previous section.

Fig.~\ref{fig:nn_residuals} shows the convergence history of various preconditioners.
The relative tolerance for solving $Ax=b$ is defined as
\begin{equation}
    \text{relative tolerance} = \frac{||Ax_0 - b||_2}{||b||_2},
    \label{eq:rel_tol}
\end{equation}
where $x_0$ is the approximate solution from the solver and $||\cdot||_2$ is the Euclidean norm.  Real and imaginary parts of the source vector $b$ are sampled from $[0,1)$ independently for all vector components. The same random source vector is used for all solves used to construct Fig.~\ref{fig:nn_residuals}. 

In all cases, neural-network preconditioners reduce the number of iterations required for convergence at any relative tolerance when compared to the unpreconditioned and even-odd preconditioned solves.
The IC preconditioner, as defined in Eq.~\eqref{eq:iceq}, is a dense triangular matrix; this approach is outperformed only by the double preconditioner constructed from the two-dense network, which improves on the IC approach while also maintaining a more sparse structure.
Comparing the various neural-network preconditioners, it is clear that---despite having the same numbers of neural-network parameters for optimization---performance is largely dictated by sparsity: dense networks outperform sparse networks, and double preconditioners outperform single preconditioners. 

For a fixed solution tolerance, we can define the improvement factor resulting from preconditioning to be
\begin{align}
    \begin{split}
        &\text{improvement factor} \equiv \\
        &~~~~~~~\frac{\text{\# CG iterations for the unprecond. solve}}{\text{\# CG iterations for the precond. solve}}.
        \label{eq:improve_factor}
    \end{split}
\end{align}
As a benchmark, we solve the preconditioned and unpreconditioned systems with a relative tolerance of $10^{-10}$ on $n_\text{test}$ configurations on each ensemble. For each case, $n_\text{src}=10$ random source vectors $b$ are used, with both the real and imaginary parts of each component sampled from $[0,1)$.  
The improvement factors of neural-network preconditioners range from $2.9$ for the single preconditioner constructed from the sparse network to $5.8$ for the double preconditioner constructed from the two-dense network, and all results are consistent with those of Fig.~\ref{fig:nn_residuals}. The improvement factors are also shown graphically in Fig.~\ref{fig:vol_improvement_facor} in Sec.~\ref{sec:vol}, together with the results of the volume transfer study.

CG iteration counts, as encoded in the improvement factor, however, are not a complete measure of preconditioner performance. 
In particular, the costs of applying neural-network preconditioners must also be taken into consideration.
Both the single and double preconditioners are constructed from the triangular matrix $L$ in Fig.~\ref{fig:arch}. For the sparse network, $L$ has approximately half as many nonzero elements as the Dirac normal matrix $A = D^\dagger D$. Naive operation counting thus suggests that applying $L$ and $L^\dagger$ sequentially should carry the same computational cost as applying $A$ once.  Assuming the cost of operator applications dominates, this suggests that an iteration of single-preconditioned CG costs twice as much as unpreconditioned CG, and double-preconditioned three times as much; that is, despite the significant reduction in the number of CG iterations for convergence, the number of operations required to apply preconditioners made by the two-dense networks would be much larger than for those made by the sparse network.

To give a quantitative comparison of the numbers of complex operations required to use different preconditioners, Tab.~\ref{tab:nn} enumerates all the neural-network preconditioners trained and the relative densities of the preconditioners and Dirac normal matrices on all ensembles.
We also include the relative densities of IC preconditioners, while the relative density of even-odd preconditioners are not shown as they can be applied analytically instead of numerically.
As expected, the relative densities of the single and double preconditioners constructed by the sparse network equal approximately $1$ and $3$, respectively.
Deviations arise from the fact that diagonal entries are retained when taking the lower triangular matrix, and the sparse convolution only preserves the sparsity in spatial and temporal dimensions and not the channel dimension. On the other hand, preconditioners constructed by the two-dense networks are much denser but their relative densities should approach a constant value at large volumes.

To evaluate the ultimate performance of various preconditioning schemes, we thus define a metric based on the number of of complex operations required to reach CG convergence. We define the relative operation advantage (ROA) at a given relative tolerance to be \begin{align}
    \begin{split}
        &\text{ROA} \equiv \\
        &~~~~~\frac{\text{\# complex operations for the unprecond. solve}}{\text{\# complex operations for the precond. solve}}.
        \label{eq:roa}
    \end{split}
\end{align}
$\text{ROA}=1$ for the unpreconditioned system; for the preconditioned system, it is given by the ratio of the improvement factor to one plus the relative preconditioner density listed in Tab.~\ref{tab:nn} for the corresponding ensemble. 

With a relative tolerance for convergence of $10^{-10}$, the ROA value for the single preconditioners constructed from the sparse network is $1.2$, which is the only of the preconditioners to achieve a value that is greater than 1. On the other hand, despite their large improvement factors, preconditioners constructed from dense networks require approximately five times more complex operations for convergence. In comparison, the naive ROA value for IC preconditioners is lower than neural-network preconditioners, because IC preconditioners are dense triangular matrices. These ROA values, however, do not take into the account the fact that the IC preconditioners are only implicitly constructed via forward substitution in practice so the total operation counts in practice will be less than those naively assumed here. On the other hand, the even-odd preconditioner achieves the best ROA value of $1.3$ despite scoring the lowest improvement factor on the nominal ensemble. The ROA values are summarized graphically in Fig.~\ref{fig:roa} in Sec.~\ref{sec:vol} in a comparison to volume-transfer results.

Although the ROA metric can be used as a guide, in practice, the precise relation between preconditioner sparsity and performance is of course even more complicated.
Beyond the details of implementation, hardware-dependent concerns such as interconnect speeds, cache sizes, and memory read/write speeds play an important role in determining performance, as the performance of most modern linear solvers in LQFT are bounded by the memory bandwidth of the hardware~\cite{Joo:2019byq}. CNN-based preconditioners with dense convolutional layers provide an opportunity to increase the arithmetic intensity---the ratio between total number of operations and the total number of bytes of data---of the solvers, thereby reducing the the total time-to-solution despite higher operation counts.
For application to theories such as lattice QCD, it will be necessary to explore the optimal method with which to implement neural-network preconditioners, taking into account these practical complications. The advantage of neural-network preconditioners then lies in the flexibility of the architecture, which can be tuned to optimize a specific metric balancing the number of iterations and the ROA value.

\subsection{Volume transfer results}
\label{sec:vol}
\begin{figure}
    \centering
    \includegraphics[width=.45\textwidth]{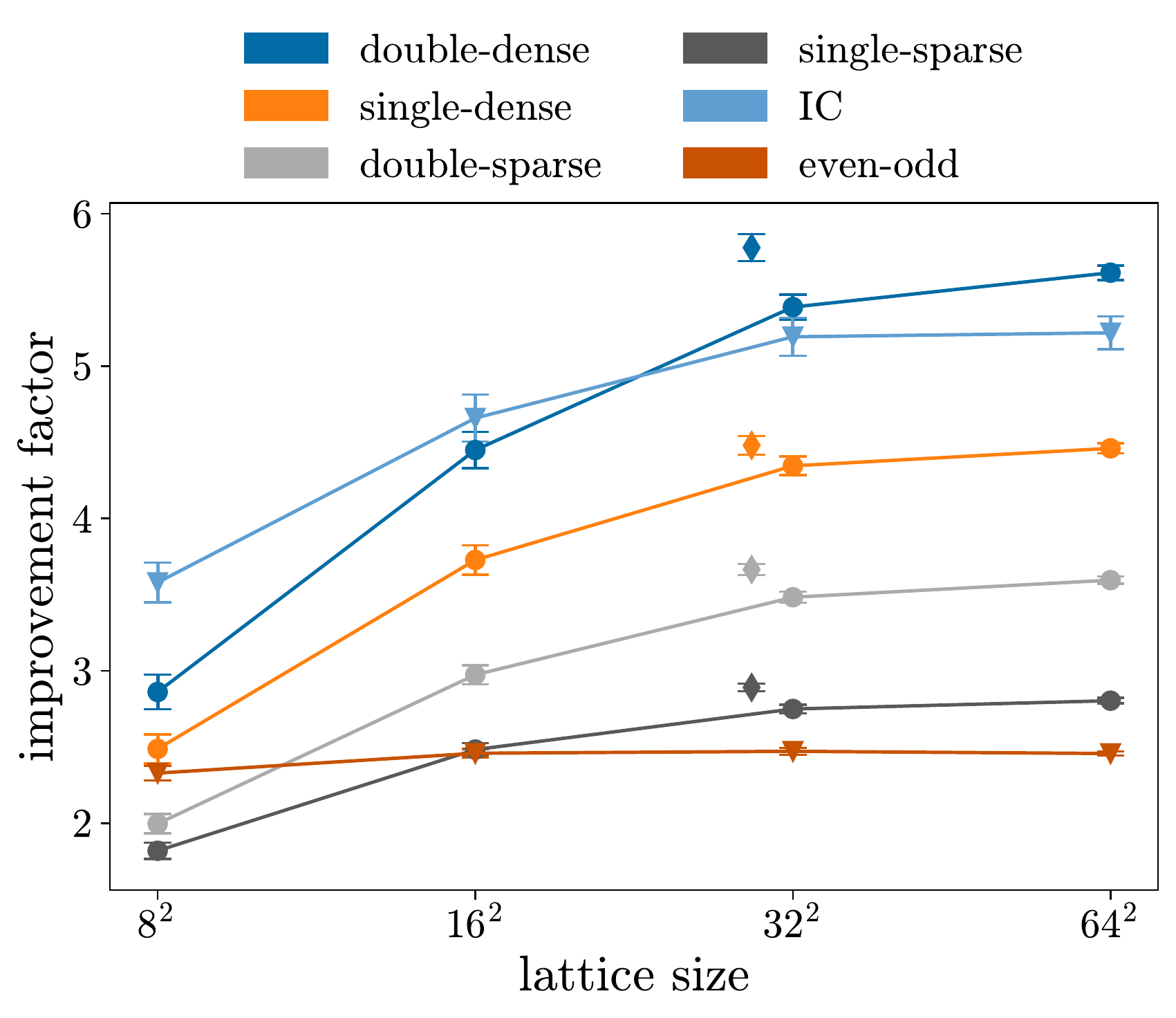}
    \caption{Improvement factors (Eq.~\eqref{eq:improve_factor}) resulting from the various preconditions applied to each of the ensembles listed in Tab.~\ref{tab:ensemble}. 
    The relative tolerance for convergence is $10^{-10}$.
    Three sets of data points are shown: the diamonds are results of neural-network preconditioners trained on the $32^2$ ensemble, the circles connected by lines are results of neural-network preconditioners trained on the $8^2$ ensemble for the volume transfer study, and the triangles connected by lines are results of IC and even-odd preconditioners. 
    The diamond markers are offset slightly on the horizontal axis for clarity. The error bars show the standard deviations of the solves.
    Legend labelling is as in Fig.~\ref{fig:nn_residuals}.}
    \label{fig:vol_improvement_facor}
\end{figure}

This section describes the results of a volume transfer study; neural networks are first trained on the $8^2$ lattice ensemble, then directly applied to larger-volume configurations to construct preconditioners without any re-optimization. Fig.~\ref{fig:vol_improvement_facor} shows the neural-network preconditioner improvement factors defined in Eq.~\eqref{eq:improve_factor} on all ensembles listed in Tab.~\ref{tab:ensemble}.
A relative tolerance of $10^{-10}$ is used as the stopping criteria for the solves, with random source vectors for the real and complex components sampled from $[0,1)$ on all lattice sites.

On the smallest $8^2$ lattice volume which we use to train the network for this study, the neural network preconditioners fail to improve on the benchmark IC preconditioners and only the neural-network preconditioners constructed by the two-dense network outperform the even-odd preconditioners. However, on lattice volumes greater than $16^2$, all neural-network preconditioners outperform the even-odd preconditioners, while the double preconditioners constructed by the two-dense network also outperform the IC preconditioners. Again, we observe that the relative densities of neural-network preconditioners  are positively correlated with their improvement factors.

\begin{figure}
    \centering
    \includegraphics[width=.45\textwidth]{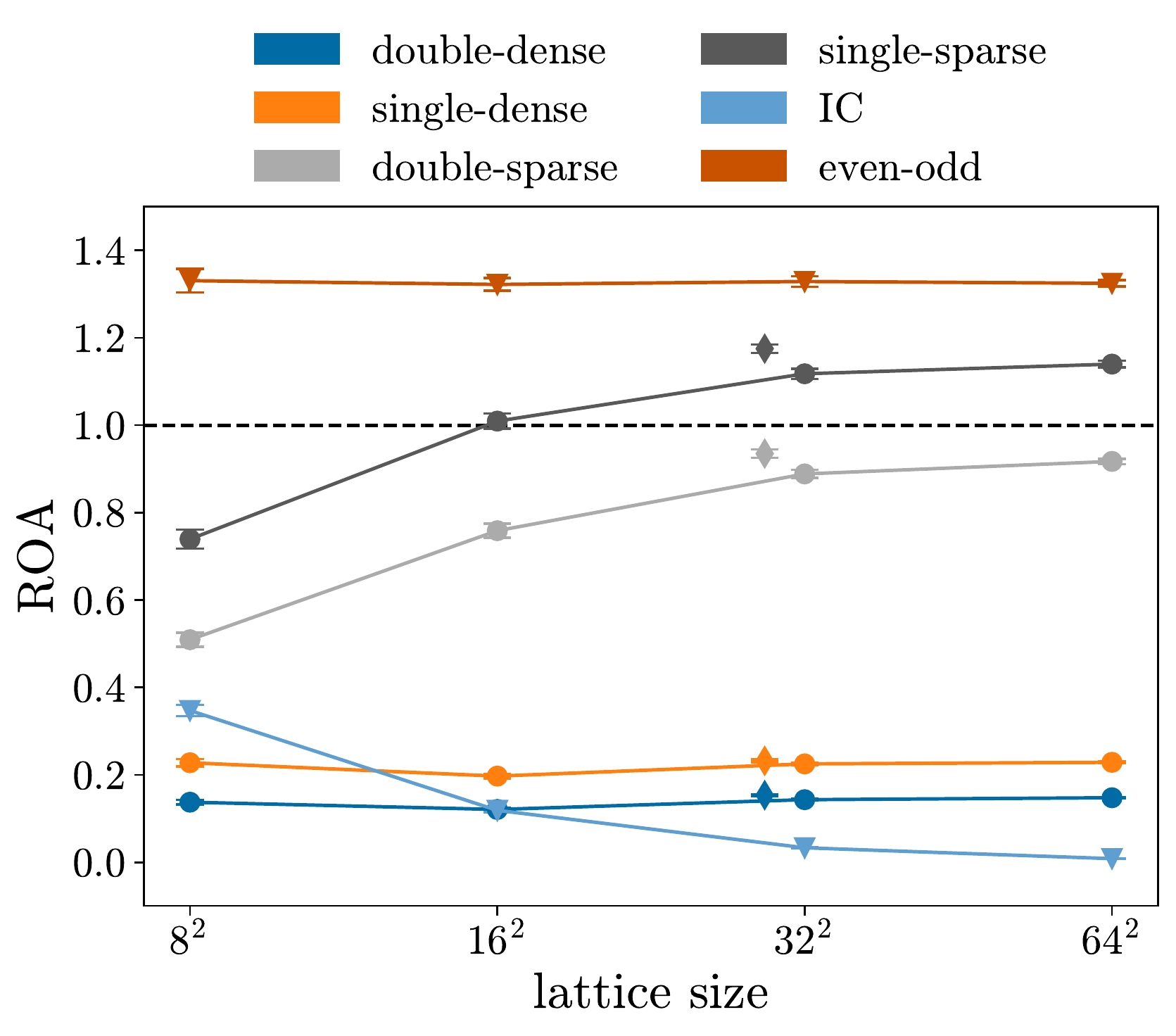}
    \caption{The preconditioner ROA values (Eq.~\eqref{eq:roa}) on ensembles listed in Tab.~\ref{tab:ensemble}. 
    The relative tolerance for convergence is $10^{-10}$.
    $\text{ROA}=1$ for  unpreconditioned systems is indicated for reference.
    Marker style is as in Fig.~\ref{fig:vol_improvement_facor} and legend labelling is as in Fig.~\ref{fig:nn_residuals}.}
    \label{fig:roa}
\end{figure}

Similarly, 
Fig.~\ref{fig:roa} shows the ROA values of neural-network preconditioners on all ensembles, evaluated with $10^{-10}$ relative tolerance.
In all cases, the ROA value is negatively correlated with the relative density of the preconditioner as shown in Tab.~\ref{tab:nn}.
The single preconditioner on the sparse network is again the only neural-network preconditioner that achieves $\text{ROA} > 1$ on lattice volumes greater than $16^2$. The even-odd preconditioner remains the most efficient preconditioner with $\text{ROA}\approx 1.4$ on all volumes.
These findings are consistent with the results of Sec.~\ref{sec:nominal_results}.

Most noticeably, even though the network is trained on a small, $8^2$ lattice with a heavy pion mass ($m_\pi = 0.47(1)$ in lattice units derived from a one-state fit to the correlation function), when we apply the network to precondition the nominal ensemble with a much lighter pion mass $(m_\pi = 0.115(10))$, its performance is similar to that of the network trained directly on the nominal ensemble as shown in Fig.~\ref{fig:vol_improvement_facor} and Fig.~\ref{fig:roa}.
In our particular implementation, it is approximately $20$ to $30$ times cheaper to train a network on the $8^2$ ensemble than on the $32^2$ ensemble, while the improvement factors and ROA values degrade minimally.
This result is particularly encouraging for future applications to lattice QCD in four dimensions, where such a volume-transfer procedure will be essential to mitigating high training costs on large lattice volumes. 

\begin{table*}[ht]
    \begin{tabular}{cccccc}
    \hline\hline
   network type & precond. type      &
   \multicolumn{4}{c}{$\frac{\text{preconditioner density}}{D^\dagger D~\text{density}}$}
    \\
    &&$8^2$&$16^2$&$32^2$&$64^2$
    \\
    \hline
     sparse & single & $1.46$&$1.46$&$1.46$& $1.46$\\
     sparse & double & $2.92$& $2.92$& $2.92$& $2.92$\\
     two-dense & single & $9.92$& $17.9$& $18.3$&$18.5$\\
     two-dense & double & $19.8$& $35.8$& $36.6$& $37.0$\\
      & IC & $9.31$& $38.2$& $155$& $625$ \\
    \hline\hline
    \end{tabular}
    \caption{Ratios of preconditioner densities to the Wilson-Dirac normal matrix density for all ensembles listed in Tab.~\ref{tab:ensemble}.}
    \label{tab:nn}
\end{table*}

\section{Discussion and Conclusion}
\label{sec:conclude}

In this work, we present CNN architectures based on both sparse and dense convolutions to construct precondtioners for solving Dirac normal matrices in the Schwinger model. In particular, we show that the neural-network preconditioners reduce the number of CG iterations required for convergence by a factor between two and five on all ensembles that we consider here, outperforming both the even-odd and IC preconditioners by this metric. 
To assess the practical value of these preconditioners, we also compare the numbers of complex operations required for convergence for the unpreconditioned and preconditioned systems, as quantified by the ROA value defined in Eq.~\eqref{eq:roa}. Using this alternate metric, single preconditioners constructed by sparse networks are the only ones to achieve an advantage over unpreconditioned solves. 
However, even though other neural-network (and IC) preconditioners require more complex operations for convergence, in practice this might not be the limiting factor to the solver performance as LQFT solvers are not compute-bound on modern architectures.

Moreover, applying a network trained on an ensemble with a $8^2$ lattice volume to generate preconditioners for ensembles with the same bare parameters but larger volumes,
the performance of these preconditioners is approximately the same as that of preconditioners constructed by a network trained directly on those larger lattice volumes.
This allows an important reduction and amortization of the training costs without significant impact on preconditioner performance. 

A natural question is whether neural-network preconditioners similar to those developed here can be used to accelerate Dirac equation solves in four-dimensional theories of phenomenological interest, such as lattice QCD. Based on this investigation, it is clear that the high computational cost of training will likely be one of the biggest hurdles that will need to be overcome to achieve a performant algorithm in this setting, regardless of the network architecture used to construct the preconditioners.
The positive volume-transfer result for the Schwinger model, however, provides an example of how training costs can be heavily ameliorated by training only on small-volume ensembles, which can be partially understood as a consequence of the locality of quantum field theories~\cite{Streater:1989vi,Liu:2017man,Luscher:2017cjh}.
In particular, locality implies that for a given set of bare parameters, the preconditioner performance of the CNN architecture trained on a given lattice volume should be almost identical to that of the same CNN trained on a smaller volume, as long as the dimensions of both volumes are much larger than the correlation length.
However, the numerical volume-transfer study illustrates that the neural-network preconditioners are still effective even when the training and testing ensembles have vastly different correlation lengths, which are characterized by the differences in the lightest pseudoscalar meson masses due to finite-size corrections. Similar investigations are thus necessary for each particular theory to determine the volume-transferring properties of neural-network preconditioners.

Beyond the question of training, attaining an in-practice advantage in time-to-solution for iterative solvers in theories such as lattice QCD will require further investigation of efficient implementations.
In this work, all algorithms have been implemented using explicit matrix representations, since this is feasible for the small problem sizes used in the proof-of-principle study. An implementation that does not rely on the explicit matrix representation will certainly be needed to generalize the approach to larger volumes.
In addition, we have shown that large numbers of complex operations are needed to apply neural-network preconditioners, especially those constructed by the network with dense convolutional layers. It may nevertheless still be feasible to accelerate the solves with dense preconditioners by maximizing the arithmetic intensity and avoiding memory limitations; possible practical advantages will rely heavily on implementation details.

Finally, combining machine learning techniques with existing preconditioning techniques is another viable approach. In particular, combining even-odd preconditioning with neural-network preconditioners has the advantage of halving the sizes of linear systems and further reducing the condition numbers. However, it is difficult to naturally define a network architecture based on convolutions for this problem, since convolutions are inefficient when operating on a lattice where only the even or the odd sites are retained; a different strategy of constructing neural-network preconditioners is thus needed to precondition even-odd preconditioned systems. In addition to sequential preconditioners, hybrid methods are possible; for example, Ref.~\cite{greenfeld2019learning} proposes a neural-network parametrization of the prolongator of the multigrid algorithm that is more efficient than the the black-box algebraic multigrid method.
On the other hand,
Ref.~\cite{DBLP:journals/corr/abs-1901-10415} shows that there is a one-to-one correspondence between operations in the CNN and the geometric multigrid algorithm,
so the lessons that have been learned in implementing multigrid algorithms for lattice QCD applications could inform future neural-network architectures, and vice versa. Additionally, scaling these hybrid approaches to typical lattice QCD problem sizes with current hardware remains challenging~\cite{10.1007/978-3-030-96498-6_3}.

Clearly, significant work remains before a practical preconditioning scheme based on neural networks can be applied to key theories such as for lattice QCD. Nevertheless, the results of this work demonstrate the viability of using machine-learned preconditioner constructions applied to a structurally similar problem, and illuminate a potential pathway towards developing neural-network preconditioners for other LQFT applications.

\acknowledgments

We thank William Detmold for useful comments on the manuscript. YL is grateful for the dicussions with Andreas Kronfeld. SC, DCH, YL, and PES are supported in part by the U.S.\ Department of Energy, Office of Science, Office of Nuclear Physics, under grant Contract Number DE-SC0011090. PES is additionally supported by the National Science Foundation under EAGER grant 2035015, by the U.S.\ DOE Early Career Award DE-SC0021006, by a NEC research award, and by the Carl G and Shirley Sontheimer Research Fund. BX is supported by the MIT UROP office. The authors acknowledge the MIT SuperCloud and Lincoln Laboratory Supercomputing Center~\cite{reuther2018interactive} for providing HPC resources that have contributed to the research results reported within this paper. Numerical experiments and data analysis used \texttt{PyTorch}~\cite{NEURIPS2019_9015}, \texttt{Minkowski Engine}~\cite{choy20194d}, \texttt{NumPy}~\cite{harris2020array}, and \texttt{SciPy}~\cite{2020SciPy-NMeth}. Figures were produced using \texttt{Matplotlib}~\cite{Hunter:2007}.

\appendix

\section{Preconditioners}
\label{appdx:precond}

\subsection{Even-odd preconditioning}\label{sec:evenodd}
Even-odd, or red-black, preconditioning \cite{DeGrand:1988vx} is one of the most commonly used preconditioners in LQFT calculations. 
If the entries in a  Dirac matrix are arranged such that the even sites precede odd sites, the matrix can be written as
\begin{equation}
    D =
    \begin{pmatrix}
    D_\text{ee}&D_\text{eo}\\
    D_\text{oe}&D_\text{oo}
    \end{pmatrix}.
    \label{eq:eo_dirac}
\end{equation}
The submatrices $D_\text{ee}$ and $D_\text{oo}$ connect even to even and odd to odd sites, and $D_\text{eo}$ and $D_\text{oe}$ connect even to odd and odd to even sites. As long as $D_\text{ee}$ and $D_\text{oo}$ can be easily inverted, left and right preconditioners can be defined as
\begin{equation}
   M_L^{-1} \equiv 
   \begin{pmatrix}
   1 & -D_\text{eo}D^{-1}_\text{oo}\\
   0 & 1
   \end{pmatrix},~
   M_R^{-1} \equiv 
   \begin{pmatrix}
   1 & 0\\
   -D^{-1}_\text{oo}D_\text{oe}&1
   \end{pmatrix},
\end{equation}
so that the preconditioned Dirac matrix is block diagonal
\begin{equation}
    D^\prime = M_L^{-1} D M_R^{-1} = 
    \begin{pmatrix}
    \bar{D}_\text{ee}&0\\
    0&D_\text{oo}\\
    \end{pmatrix},
    \label{eq:eo}
\end{equation}
where $\bar{D}_\text{ee} = D_\text{ee} - D_\text{eo}D^{-1}_\text{oo}D_\text{oe}$. The CG algorithm can then be applied to the normal equation of the preconditioned matrix
\begin{equation}
A^\prime \equiv (\bar{D}_\text{ee})^\dagger\bar{D}_\text{ee}.
\label{eq:eo_normal}
\end{equation}
Note that $\bar{D}_\text{ee}$ is simply the Schur complement of the Dirac matrix in Eq.~\eqref{eq:eo_dirac}. We observe that the even-odd decomposition provides a factor of $2$$-$$3$ reduction in iteration numbers required for convergence for Dirac matrices used in this work.

\subsection{Incomplete Cholesky decomposition}
Let $A$ be a sparse HPD matrix of order $n$. The incomplete Chloesky (IC) decomposition of $A$ that we use in this work is defined as
\begin{align}
    \begin{split}
    (L_{c})_{ii} &= \sqrt{A_{ii} - \sum_{k=1}^{i-1}(L_{c})_{ik}(L_{c})^*_{ik}},\\
    (L_{c})_{ij}&= \frac{1}{(L_{c})_{jj}}\left(
    A_{ij} - \sum_{k=1}^{j-1} (L_{c})_{ik}(L_{c})^*_{jk}
    \right), \\
    &~~~~~~~~~~~~~~~~~~~~~~~~~~~~~~~~~~~
    i>j,A_{ij} \neq 0,
    \end{split}
    \label{eq:ichol}
\end{align}
where $L_c$ is a lower triangular matrix. Eq.~\eqref{eq:ichol} is similar to the Cholesky decomposition, except the IC algorithm imposes the constraint that the matrix $L_c$ has the same sparsity pattern as the lower triangular part of the matrix $A$. In the applied mathematics literature, this specific decomposition is called the IC decomposition with zero fill-ins, or IC$(0)$~\cite{booksaad}. Note that Eq.~\eqref{eq:ichol} implies that
\begin{equation}
    A_{ij} = \sum_{k=1}^{n}(L_c)_{ik}(L_c)^*_{jk}
\end{equation}
for $(i,j)$ where $A_{ij}\neq 0$, so the IC decomposition reproduces non-zero entries of $A$ exactly; however, $L_c L_c^\dagger$ is in general denser than $A$. 

To use $L_c$ as a preconditioner, we define the left and right preconditioners, $M^{-1}_L$ and $M^{-1}_R$, to be
\begin{equation}
    M^{-1}_L = (M^{-1}_R)^\dagger = L_c^{-1},
    \label{eq:iceq}
\end{equation}
where the inverses of triangular matrices could be computed with the forward substitution method.

\bibliographystyle{apsrev4-2}
\bibliography{main}
\end{document}